\documentstyle[aps]{revtex}
\def\bea{\begin{eqnarray}}
\def\eea{\end{eqnarray}}
\def\nn{\nonumber}
\def\r{\circle*{1}}
\def\cn{{\cal N}}
\begin{document}
\tolerance 50000
\title{Magnetic moment interaction in the anyon superconductor}
\author{M. Eliashvili and G. Tsitsishvili}
\address{Department of Theoretical Physics\\
A. Razmadze Institute of Mathematics\\
Tbilisi 380093, Georgia}
\twocolumn[
\maketitle
\widetext
\vspace*{-1.0truecm}
\begin{abstract}
\begin{center}
\parbox{14cm}{
Magnetic moment interaction is shown to play a defining
role in the magnetic properties of anyon superconductors.
The necessary condition for the existence of the Meissner effect
is found.}
\end{center}
\end{abstract}
]
\narrowtext

The zero-temperature Meissner effect presented in the 2+1
dimensional anyon matter provoked  considerable efforts
in order to promote the Chern-Simons gauge theory as a
hypothetical candidate for the high-$T_c$ superconductivity.

The most important points in that development are existence
of the massless pole in the current correlators [1],
cancellation of bare and induced Chern-Simons  terms [2] and
detailed calculations of effective action and thermodynamic
potential for the fermions interacting with Chern-Simons and
Maxwell fields [3,4,5,6].

As a convincing argument in favor of the superconducting nature
of the anyon system, one can use the energetic one: energy density
evaluated as a function of the Maxwell magnetic field exhibits a
cusplike structure with a minima located at the integer values
of the filling fraction. Note that this conclusion has been
reached studying the nonrelativistic model with neglected
magnetic moment interactions [5].

In the present paper we reconsider the problem with the aim to
clarify the role played by the spin interaction. The necessity
to consider such a term is prompted by the relativistic Lagrangian
\bea
L=i\hbar c\bar\Psi\gamma^\mu D_\mu\Psi-\sigma mc^2\bar\Psi\Psi,\nn
\eea
\bea
D_\mu=\partial_\mu+i(e/\hbar)(A_\mu+a_\mu),\nn
\eea
where $m>0$, $\sigma=\pm1$, and $\Psi$ is a two-component spinor.
Unlike the 3+1 dimensional case,  where the sign of the mass term
does not play any role, here $\sigma$ describes the helicity of
2+1 dimensional relativistic particle and the values $\pm1$
correspond to the different particle types, belonging to the
different representations of the 2+1 dimensional Poincare group.

We consider the case where $A_0=a_0=0$, while $A_k$ and $a_k$
correspond to the homogeneous Maxwell $(B)$ and Chern-Simons $(b)$
magnetic fields, respectively. Then, the relativistic Hamiltonian
and the corresponding non-relativistic one are given by
\bea
H=i\hbar c\Psi^\dagger\gamma^0\gamma_kD_k\Psi
+\sigma mc^2\Psi^\dagger\gamma^0\Psi,
\eea
\bea
H^{\rm nr}=\frac{\hbar^2}{2m}|D_k\psi|^2
-\sigma\frac{e\hbar}{2m}(B+b)\psi^\dagger\psi,
\eea
where $\psi$ is the one-component nonrelativistic matter field,
and the magnetic moment interaction in (2) includes both Maxwell
and Chern-Simons contributions. In the previous considerations [5]
the second term in (2) was completely neglected ($\sigma=0$).
Evidently, this cannot be done in (1). The effects related
to magnetic moment interactions can be included dealing either
with (1) or (2).

In the relativistic theory the covariant coupling leads to the
magnetic interaction of $(B+b)$-type only, which in the
nonrelativistic limit is reduced to (2). However, in the
nonrelativistic treatment one can admit the extra contributions
of the Chern-Simons magnetic fields:
$H^{\rm nr}\to H^{\rm nr}+\lambda b\psi^\dagger\psi$.
Taking into account that $b$ is fixed by the net particle density,
and therefore is a constant, one concludes that the $\lambda$ term
simply defines the energy scale and does not lead to any new
effects in magnetic properties of the system.

To be complete, we consider the relativistic version and imply
the normal ordering of the fermion operators in Hamiltonian
and particle number operator. In this consideration Hamiltonian
(1) becomes positively defined, and the planar density
of relativistic thermodynamic potential looks as follows:
\bea
\Omega=\frac{k_BT}{2\pi\ell^2}\left\{\frac{1+\sigma\epsilon}{2}
\ln(1-\rho_0)+\sum_{n=1}^{\infty}\ln(1-\rho_n)\right\}+\nn
\eea
\bea
+\frac{k_BT}{2\pi\ell^2}\left\{\frac{1-\sigma\epsilon}{2}
\ln(1-\bar\rho_0)+\sum_{n=1}^{\infty}
\ln(1-\bar\rho_n)\right\},
\eea
where $\rho_n$ and $\bar\rho_n$ are the Fermi distribution
functions for particles and antiparticles, respectively,
\bea
\rho_n=\left[1+e^{(E_n-\mu)/k_BT}\right]^{-1},\nn
\eea
\bea
\bar\rho_n=\left[1+e^{(E_n+\mu)/k_BT}\right]^{-1}.\nn
\eea
Relativistic energies $E_n$ are given by
\bea
E_n=mc^2\left[1+2(\lambda^2/\ell^2)n\right]^{1/2},
\eea
where $\lambda=\hbar/mc$ is the Compton wavelength,
while the magnetic length $\ell$ and the parameter
$\epsilon$ are defined by
\bea
\frac{1}{\ell^2}=\frac{1}{\hbar}|eB+eb|,
\eea
\bea
\epsilon={\rm sgn}(eB+eb).\nn
\eea

The fact that the relativistic thermodynamic potential does not
contain the $n=0$ terms corresponding to particles
(when $\sigma=-\epsilon$) or antiparticles
(when $\sigma=+\epsilon$) is the consequence
of the absence of these modes in the spectrum
of the relativistic one-particle Hamiltonian.

The system can be described by the Helmholtz free energy
\bea
F(B+b,T,\cn)=\Omega(B+b,T,\mu)+\mu\cn,
\eea
where the chemical potential $\mu=\mu(B+b,T,\cn)$
should be found out from the equation
\bea
\cn=-\frac{\partial\Omega}{\partial\mu}\equiv n_e-n_{\bar e}.\nn
\eea
Here $n_e$ and $n_{\bar e}$ are the average densities
of particles and antiparticles, respectively,
\bea
n_e=\frac{1}{2\pi\ell^2}\left\{\frac{1+\sigma\epsilon}{2}\rho_0+
\sum_{n=1}^{\infty}\rho_n\right\},\nn
\eea
\bea
n_{\bar e}=\frac{1}{2\pi\ell^2}\left\{\frac{1-\sigma\epsilon}{2}
\bar\rho_0+\sum_{n=1}^{\infty}\bar\rho_n\right\}.\nn
\eea

Using the usual definition of the filling fraction
$(\nu=2\pi\ell^2\cn)$ and of magnetic length (5) we get
\bea
\nu=\frac{2\pi\hbar}{\epsilon e}\frac{\cn}{B+b}
\eea
implying that for a given $\cn$ the filling fraction can be
used instead of $B+b$ as a one of the independent arguments
in the free energy (6). In terms of the distribution functions
it appears as
\bea
\nu=\frac{1+\sigma\epsilon}{2}\rho_0
-\frac{1-\sigma\epsilon}{2}\bar\rho_0
+\sum_{n=1}^{\infty}(\rho_n-\bar\rho_n).\nn
\eea

The mechanism leading to the Meissner effect is based on
the assumption that at some value $\nu=\nu_0$ the free energy
$F(\nu,T,\cn)$ possesses a minimum. In that case, choosing
the corresponding value of the Chern-Simons magnetic field to be
\bea
b=\frac{2\pi\hbar}{\epsilon e}\frac{\cn}{\nu_0},\nn
\eea
we observe that this minimum is achieved at $B=0$, while any small
$B\ne0$ costs some positive energy.

The free energy calculated for the nonrelativistic Hamiltonian (2)
with completely dropped magnetic moment interaction possesses
the local minima at the integer values of filling fraction [5],
where the Meissner effect just takes the place.

To simplify the present account, note that due to
$\rho_n(\mu)=\bar\rho_n(-\mu)$ we have
$\Omega(\sigma,\mu)=\Omega(-\sigma,-\mu)$ reflecting
the invariance of the relativistic thermodynamic potential (3)
under the interchange of particles and antiparticles.
Correspondingly, the free energy (6) is invariant under
$\{\sigma\to-\sigma$, $n_e\rightleftharpoons n_{\bar e}\}$.
Therefore, without loss of generality we can deal with $\mu>0$.
Consequently, due to $E_n+\mu>mc^2$ one has
$\bar\rho_n<\exp(-mc^2/k_BT)$ and assuming
\bea
\frac{mc^2}{k_BT}>>1,
\eea
we get the antiparticle contributions to be effectively zero
$(\bar\rho_n=0)$, implying $n_{\bar e}=0$ and $\cn=n_e$
(note that for $m$ being of order of an electron mass $m\sim m_e$
and $T\sim100$K we have $mc^2/k_BT\sim10^7$).

In what follows we present the zero-temperature analysis and
comment on the changes appearing at finite temperatures in the end.

Due to $\bar\rho_n=0$ we have $\nu\geq0$, and presenting
the filling fraction as $\nu=N+\theta$, where $0\leq\theta\leq1$,
$N=0,1,2,\ldots$ we write down the zero-temperature values of
the particle distribution functions as
\bea
\rho_n=\left\{\begin{array}{c}\vspace*{3mm}
1\hspace*{5mm}{\rm if}\hspace*{5mm}n<N+(1/2)(1-\sigma\epsilon)\\
\vspace*{3mm}\theta\hspace*{5mm}{\rm if}\hspace*{5mm}n=N+(1/2)(1-\sigma\epsilon)\\
0\hspace*{5mm}{\rm if}\hspace*{5mm}n>N+(1/2)(1-\sigma\epsilon).\end{array}\right.
\eea

The free energy at $T=0$ coincides with the internal energy
$U=(1/2\pi\ell^2)\sum\rho_nE_n$.
For $0\leq\nu\leq1$ $(N=0)$ one gets
\bea
F=mc^2n_e
\left[1+2\pi n_e\lambda^2(1-\sigma\epsilon)\nu^{-1}\right]^{1/2}.
\eea

To carry out the analysis for $\nu\geq1$ we assume
\bea
n_e\lambda^2<<1
\eea
(for $m\sim m_e$ and the typical value [5] $n_e\sim10^{14}cm^{-2}$,
$n_e\lambda^2\sim10^{-7}$). Further, summation over the particle
contributions has the upper bound defined in (9). Up to this value
of $n$ due to (11) we can use
\bea
(\lambda^2/\ell^2)n=2\pi n_e\lambda^2(n/\nu)<<1,
\eea
and the corresponding relativistic energies (4) are effectively
reduced to the nonrelativistic ones
\bea
E_n=mc^2+\frac{\hbar^2}{m\ell^2}n,\nn
\eea
leading to the following expression of the free energy:
\bea
F=mc^2n_e+\frac{\pi\hbar^2 n_e^2}{m}
\left[1+\frac{\theta(1-\theta)}{(N+\theta)^2}-
\frac{\sigma\epsilon}{N+\theta}\right].
\eea

Note that for $\sigma=+\epsilon$ the expression (13) with $N=0$
coincides with (10). For $\sigma=-\epsilon$, the reduction of
(10) to (13) with $N=0$ becomes invalid for $\nu\sim0$ where
the condition (12) cannot be justified. However, here we can
exclude this region from our consideration, since
the Meissner effect is expected to take the place near the
integer values of $\nu$.

The second term in (13) is the nonrelativistic free energy,
which could be obtained if one starts with Hamiltonian (2).
It should be pointed out that the nonrelativistic expression
(13) was obtained from the relativistic considerations using
the assumptions $mc^2>>k_BT$ (8) and $n_e<<\lambda^{-2}$ (11),
but not a direct nonrelativistic limit $(c\to\infty)$.

Omitting the relativistic contribution $mc^2n_e$, we present
$F(\nu)$ for the different values of $\sigma$ in figure 1.
The case $\sigma=0$ corresponds to the non-relativistic
Hamiltonian (2) with completely neglected magnetic moment
interaction, but not to the massless relativistic fermions.
As one can see, the local minima occurring for $\sigma=0$ are
lost for $\sigma=\pm\epsilon\ne0$. Therefore, the relativistic
fermionic system (1) as well as the nonrelativistic one (2)
with preserved magnetic moment interaction does not exhibit
the Meissner effect.
\vspace*{8mm}
\begin{center}
\begin{picture}(336,294)
\put(0,24){\line(1,0){336}} \multiput(96,20)(72,0){3}{\line(0,1){8}}
\put(24,0){\line(0,1){294}} \multiput(20,40)(0,100){3}{\line(1,0){8}}
\put(93,4){$1$} \put(165,4){$2$} \put(237,4){$3$} \put(307,4){$\nu$}
\put(6,37){$0$} \put(6,137){$1$} \put(6,237){$2$}
\put(50,280){$(m/\pi\hbar^2n_e^2)F$}

\multiput(33,40)(3,0){22}{\r}
\put(99,49){\r}   \put(102,57){\r}  \put(105,64){\r}  \put(108,69){\r}
\put(111,74){\r}  \put(114,78){\r}  \put(117,82){\r}  \put(120,85){\r}
\put(123,88){\r}  \put(126,90){\r}  \put(129,92){\r}  \put(132,93){\r}
\put(135,95){\r}  \put(138,96){\r}  \put(141,97){\r}  \put(144,98){\r}
\put(147,98){\r}  \put(150,99){\r}  \put(153,99){\r}  \put(156,100){\r}
\put(159,100){\r} \put(162,100){\r} \put(165,100){\r} \put(168,100){\r}
\put(171,102){\r} \put(174,105){\r} \put(177,106){\r} \put(180,108){\r}
\put(183,110){\r} \put(186,111){\r} \put(189,112){\r} \put(192,113){\r}
\put(195,114){\r} \put(198,115){\r} \put(201,116){\r} \put(204,117){\r}
\put(207,117){\r} \put(210,118){\r} \put(213,118){\r} \put(216,119){\r}
\put(219,119){\r} \put(222,119){\r} \put(225,120){\r} \put(228,120){\r}
\put(231,120){\r} \put(234,120){\r} \put(237,120){\r} \put(240,120){\r}
\put(243,121){\r} \put(246,122){\r} \put(249,123){\r} \put(252,124){\r}
\put(255,125){\r} \put(258,125){\r} \put(261,126){\r} \put(264,126){\r}
\put(280,126){$_{\sigma=+\epsilon}$}

\put(90,261){\r}  \put(93,250){\r}
\put(96,240){\r}  \put(99,240){\r}  \put(102,239){\r} \put(105,237){\r}
\put(108,235){\r} \put(111,233){\r} \put(114,230){\r} \put(117,228){\r}
\put(120,225){\r} \put(123,222){\r} \put(126,219){\r} \put(129,216){\r}
\put(132,213){\r} \put(135,210){\r} \put(138,207){\r} \put(141,204){\r}
\put(144,202){\r} \put(147,199){\r} \put(150,196){\r} \put(153,193){\r}
\put(156,190){\r} \put(159,188){\r} \put(162,185){\r} \put(165,183){\r}
\put(168,180){\r} \put(171,180){\r} \put(174,180){\r} \put(177,179){\r}
\put(180,179){\r} \put(183,178){\r} \put(186,178){\r} \put(189,177){\r}
\put(192,176){\r} \put(195,176){\r} \put(198,175){\r} \put(201,174){\r}
\put(204,173){\r} \put(207,172){\r} \put(210,171){\r} \put(213,170){\r}
\put(216,169){\r} \put(219,168){\r} \put(222,167){\r} \put(225,166){\r}
\put(228,164){\r} \put(231,163){\r} \put(234,162){\r} \put(237,161){\r}
\put(240,160){\r} \put(243,160){\r} \put(246,160){\r} \put(249,159){\r}
\put(252,159){\r} \put(255,158){\r} \put(258,158){\r} \put(261,157){\r}
\put(264,157){\r} \put(280,158){$_{\sigma=-\epsilon}$}

\put(60,261){\r}  \put(63,241){\r}
\put(66,226){\r}  \put(69,212){\r}  \put(72,200){\r}  \put(75,189){\r}
\put(78,180){\r}  \put(81,172){\r}  \put(84,164){\r}  \put(87,157){\r}
\put(90,151){\r}  \put(93,145){\r}  \put(96,140){\r}  \put(99,144){\r}
\put(102,148){\r} \put(105,150){\r} \put(108,152){\r} \put(111,154){\r}
\put(114,155){\r} \put(117,156){\r} \put(120,156){\r} \put(123,156){\r}
\put(126,156){\r} \put(129,155){\r} \put(132,154){\r} \put(135,153){\r}
\put(138,152){\r} \put(141,151){\r} \put(144,150){\r} \put(147,148){\r}
\put(150,147){\r} \put(153,146){\r} \put(156,145){\r} \put(159,144){\r}
\put(162,142){\r} \put(165,141){\r} \put(168,140){\r} \put(171,141){\r}
\put(174,142){\r} \put(177,143){\r} \put(180,144){\r} \put(183,144){\r}
\put(186,145){\r} \put(189,145){\r} \put(192,146){\r} \put(195,146){\r}
\put(198,147){\r} \put(201,147){\r} \put(204,147){\r} \put(207,146){\r}
\put(210,146){\r} \put(213,146){\r} \put(216,145){\r} \put(219,145){\r}
\put(222,144){\r} \put(225,144){\r} \put(228,143){\r} \put(231,143){\r}
\put(234,142){\r} \put(237,141){\r} \put(240,140){\r} \put(243,141){\r}
\put(246,142){\r} \put(249,143){\r} \put(252,143){\r} \put(255,144){\r}
\put(258,144){\r} \put(261,145){\r} \put(264,145){\r}
\put(280,142){$_{\sigma=0}$}
\end{picture}\vskip2mm
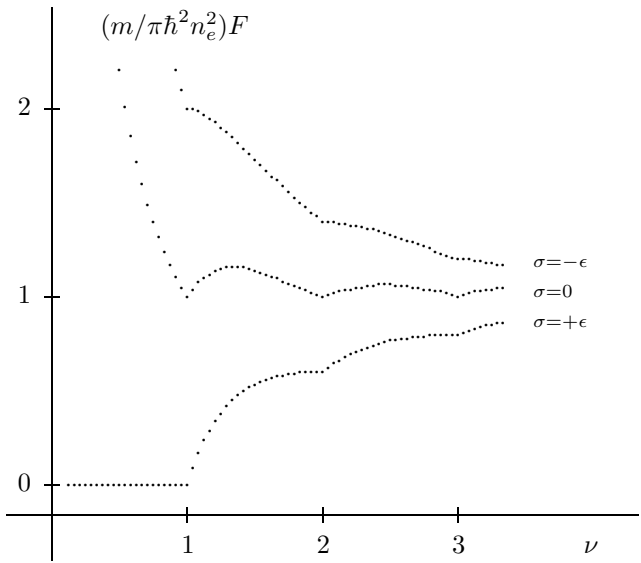
Figure 1. {\small Free energy of the one-component system.}
\end{center}
\vspace*{8mm}

Consider now the system containing two types of fermions with
opposite values of $\sigma$. In that case the system
is the combination of two subsystems with $\sigma=\pm\epsilon$
and the corresponding quantities will be distinguished by the
uppercase indices $\pm$. Now, the relation (7)
(with $n^\pm_{\bar e}=0$) should be rewritten as
\bea
\nu=\frac{2\pi\hbar}{\epsilon e}\frac{n^+_e+n^-_e}{B+b}=\nu^++\nu^-,
\eea
where the partial filling fractions appear as
\bea
\nu^\pm=\frac{1\pm1}{2}\rho^\pm_0+\sum_{n=1}^{\infty}\rho^\pm_n.\nn
\eea

The system is assumed to be in contact with a particle reservoir
which keeps the total particle density $n^{}_e=n^+_e+n^-_e$
fixed and guarantees the chemical equilibrium, {\it i.e.} converts
the particles of one type into another and {\it vice versa}
if energetically favorable. Taking into account the energy
spectrum with the aforementioned asymmetry of $n=0$ modes,
one can easily derive the conditions for the chemical equilibrium
between the two subsystems.

If the density $n_e$ is small enough $(2\pi\ell^2n_e\leq1)$,
then all particles will be disposed on the level with the lowest
energy $E_0$, and we have
\bea
\begin{array}{c}\vspace*{3mm}0\leq\nu^+\leq1,\\ \nu^-=0.\end{array}
\eea
The corresponding zero-temperature values of
the distribution functions are
\bea
\rho^+_0=\nu^+,\hspace*{15mm}\rho^+_{n>0}=\rho^-_n=0.\nn
\eea

If $2\pi\ell^2n_e>1$, the levels with higher energies have to be
also occupied, and the equilibrium condition is given by
\bea
\begin{array}{c}\vspace*{3mm}N+1\leq\nu^+\leq N+2,\\ N\leq\nu^-\leq N+1,\end{array}
\eea
where $N=0,1,\ldots$ . Using the parametrizations
$\nu^+=N+1+\theta^+$ and $\nu^-=N+\theta^-$ with
$0\leq\theta^\pm\leq1$ we have
\bea
\rho^\pm_n=\left\{\begin{array}{lll}\vspace*{3mm}
1&{\rm if}&n<N+1\\\vspace*{3mm}
\theta^\pm&{\rm if}&n=N+1\\
0&{\rm if}&n>N+1.\end{array}\right.\nn
\eea

The Helmholtz free energy $F=F^++F^-$ at $T=0$ takes the value
$mc^2n_e$ in the case (15), while for (16) it is given by
\bea
F=mc^2n_e+\frac{2\pi\hbar^2n_e^2}{m}
\frac{(N+1)(N+2\theta)}{(2N+1+2\theta)^2}
\eea
and, as we see, depends on $2\theta\equiv\theta^++\theta^-$,
but not on $\theta^+-\theta^-$. Consequently, the free energy
can be considered as a function of $\nu=\nu^++\nu^-$ and the
corresponding curve is presented in figure 2 (we have
omitted the relativistic contribution $mc^2n_e$).
\vspace*{8mm}
\begin{center}
\begin{picture}(336,144)
\put(0,24){\line(1,0){336}} \put(24,0){\line(0,1){144}} \put(6,37){$0$}
\multiput(60,20)(36,0){7}{\line(0,1){8}} \put(20,40){\line(1,0){8}}
\put(57,4){$1$} \put(93,4){$2$} \put(129,4){$3$} \put(165,4){$4$}
\put(201,4){$5$} \put(237,4){$6$} \put(272,4){$7$} \put(307,4){$\nu$}
\put(6,130){$F$}

\multiput(32,40)(4,0){8}{\r}
\put(61,50){\r}   \put(62,58){\r}   \put(63,66){\r}   \put(64,73){\r}
\put(65,80){\r}   \put(66,87){\r}   \put(67,92){\r}   \put(68,97){\r}
\put(69,101){\r}  \put(70,106){\r}  \put(71,109){\r}  \put(72,113){\r}
\put(73,117){\r}  \put(74,121){\r}  \put(76,125){\r}  \put(77,128){\r}
\put(79,131){\r}  \put(81,134){\r}  \put(84,136){\r}  \put(87,138){\r}
\put(90,139){\r}  \put(93,140){\r}  \put(96,140){\r}  \put(99,140){\r}
\put(102,139){\r} \put(105,138){\r} \put(108,136){\r} \put(111,133){\r}
\put(114,130){\r} \put(116,127){\r} \put(118,124){\r} \put(120,121){\r}
\put(122,117){\r} \put(124,113){\r} \put(126,108){\r} \put(128,103){\r}
\put(130,97){\r}  \put(132,91){\r}  \put(135,97){\r}  \put(137,103){\r}
\put(139,108){\r} \put(141,113){\r} \put(143,117){\r} \put(145,121){\r}
\put(147,125){\r} \put(150,129){\r} \put(153,133){\r} \put(156,136){\r}
\put(159,138){\r} \put(162,139){\r} \put(165,140){\r} \put(168,140){\r}
\put(171,140){\r} \put(174,140){\r} \put(177,139){\r} \put(180,138){\r}
\put(183,136){\r} \put(186,134){\r} \put(189,132){\r} \put(192,129){\r}
\put(196,125){\r} \put(200,120){\r} \put(204,114){\r} \put(207,119){\r}
\put(210,123){\r} \put(213,126){\r} \put(216,129){\r} \put(219,132){\r}
\put(222,134){\r} \put(225,136){\r} \put(229,138){\r} \put(233,139){\r}
\put(237,140){\r} \put(241,140){\r} \put(245,140){\r} \put(249,140){\r}
\put(253,139){\r} \put(257,138){\r} \put(261,136){\r} \put(265,134){\r}
\put(269,131){\r} \put(273,128){\r} \put(276,125){\r} \put(280,128){\r}
\put(284,131){\r} \put(288,134){\r} \put(292,136){\r} \put(296,138){\r}
\put(300,139){\r} \put(304,140){\r} \put(308,140){\r} \put(312,140){\r}
\end{picture}\vskip2mm
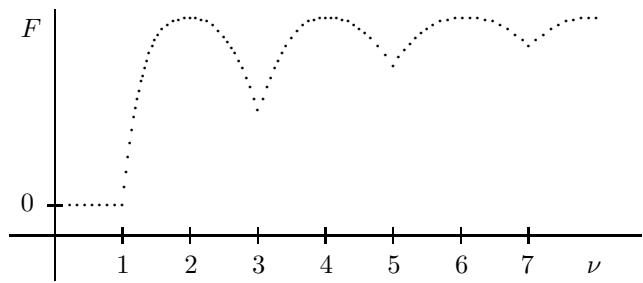
Figure 2. {\small Free energy of the compound system.}
\end{center}

As one can see, the local minima are restored for
$\nu=2K+1=3,5,\ldots$ , and the values corresponding to these
minima are given by
\bea
F_{\rm min}=\frac{\pi\hbar^2n_e^2}{2m}
\left[1-\frac{1}{(2K+1)^2}\right].\nn
\eea

In order to comment on the basic changes brought by the finite
temperature corrections, let us first point out the main
features of the zero-temperature case presented in figure 1.
\begin{itemize}
\item[(a)]
For $\sigma=0$, the magnetization $M=-\partial F/\partial B$
changes the sign from $"+"$ to $"-"$ at the integer values of
$\nu$. This implies the susceptibility
$\chi=-\partial M/\partial B$ to be positive $(\chi>0)$,
confirming the existence of the Meissner effect.
\item[(b)]
The curve $\sigma=0$ is not smooth at $\nu=integer$, {\it i.e.}
the magnetization undergoes the discontinuity, which on its term
means that the susceptibility takes the infinite magnitude,
implying that the Meissner effect is perfect.
\item[(c)]
Taking into account the magnetic moment interaction
$(\sigma=\pm\epsilon)$, the magnetization does not change
the sign at $\nu=integer$, and one loses the perfect Meissner
effect, which can be restored via the duplication of the fermionic
degrees of freedom.
\end{itemize}

The nonanalyticity of zero-temperature free energy
at the integer values of $\nu$ mentioned in (b) is
the result of the nonanalytical behaviour of
zero-temperature chemical potential, which is a steplike
function at the integer values of $\nu$.

At finite temperatures the chemical potential, and consequently
the free energy, become smooth with respect to $\nu$.
At $T\ne0$ the magnetization corresponding to $\sigma=0$ still
changes the sign, from $"+"$ to $"-"$ at $\nu=integer$, and the
susceptibility is still positive leading to the Meissner effect,
which however is partial since the susceptibility turns out to be
finite at $T\ne0$. Accounting for the magnetic moment interactions
$(\sigma=\pm\epsilon)$ at $T\ne0$, one loses the partial
Meissner effect, which can be restored in the framework of a
duplicated system. So, at $T\ne0$ one observes that only point
(b) is changed.

Thus the final conclusion can be formulated as follows:
the circumstance, whether the temperature is zero or finite,
defines whether the Meissner effect is perfect or partial,
while the magnetic moment interactions determine whether
the Meissner effect (perfect or partial) can exist at all.

This work was supported by the Georgian Academy of Sciences,
under Grant No. 1.4.

\end{document}